**Radiation-Triggered Superfluorescent Scintillation in Quantum-Ordered Perovskite Nanocrystal Superlattices**

Matteo L. Zaffalon[1,2], Andrea Fratelli[1,3], Taras Sekh[4], Emanuele Mazzola[2,5], Francesco Carulli[1], Francesco Bruni[1,2], Maryna Bodnarchuk[4], Francesco Meinardi[1], Luca Gironi[2,5], Maksym V. Kovalenko[4]*, Sergio Brovelli[1,2]*

[1]*Dipartimento di Scienza dei Materiali, Università degli Studi di Milano Bicocca, via R. Cozzi 55, Milano, Italy.*
[2]*INFN - Sezione di Milano - Bicocca, Milano, Italy.*
[3]*Nanochemistry, Istituto Italiano di Tecnologia, via Morego 30, Genova, Italy.*
[4]*Department of Chemistry and Applied Bioscience, ETH Zürich, Zürich, Switzerland.*
*Laboratory for Thin Films and Photovoltaics and Laboratory for Transport at Nanoscale Interfaces, Empa – Swiss Federal Laboratories for Materials Science and Technology, Dübendorf, Switzerland.*
[5]*Dipartimento di Fisica, Università degli Studi di Milano-Bicocca, Piazza della Scienza, Milano, Italy.*

*Corresponding authors: sergio.brovelli@unimib.it, mvkovalenko@ethz.ch*

**Abstract**

Superfluorescence, a cooperative emission phenomenon arising from the coherent coupling of excited dipoles, has historically been observed under optical excitation in carefully engineered quantum systems. Here, we report the first observation of superfluorescence triggered by ionizing radiation in lead-halide perovskite nanocrystal (NC) superlattices. Using $CsPbBr_3$ NC superlattices with long-range structural and electronic order, we demonstrate that secondary electrons generated by high-energy photons can induce efficient cooperative emission bursts characteristic of superfluorescence with unprecedented scintillation lifetime of ~40 ps, thereby introducing a new class of coherent scintillating metamaterials. Side-by-side optical and scintillation measurements reveal a direct analogy between ionizing and intense optical excitation, both leading to high excitonic densities that result in superfluorescent emission, even at mild, technologically accessible cryogenic temperatures. The discovery that incoherent, stochastic ionization cascades can seed coherent many-body optical responses with radiatively accelerated luminescence and large Stokes shifts establishes a pathway toward ultrafast, reabsorption-free, quantum-ordered nanotechnological scintillators, paving the way for the future development of radiation detectors based on quantum technologies for advanced radiation detection applications.

## Main

Ultrafast detection of ionising radiation underpins modern medical imaging[1], high-energy physics (HEP)[2-4], nuclear security[5] and space instrumentation[6]. Despite more than half a century of optimisation in crystal growth, dopant chemistry and polymer design, virtually every commercial scintillator still channels the deposited energy into incoherent, stochastic recombination of charge carriers. In both high-Z inorganic crystals and organic plastics, the light yield (LY, defined as the number of emitted photons per unit of deposited energy), timing profile, and photon statistics are ultimately determined by the stochastic capture and recombination of localized charge carriers via disorder- or phonon-mediated pathways. These fundamental limitations impair the time-of-flight resolution in positron emission tomography (PET) scanners and the vertex reconstruction accuracy in HEP detectors. They also prevent the implementation of quantum-based readout strategies, such as those currently employed in quantum communication[7] and sensing[8], as well as the deliberate tuning of scintillation properties by engineering many-body electronic states to enhance radiative rates, shape emission waveforms, shift spectra away from self-absorption, or control photon-number statistics.

In principle, collective coherent phenomena like superfluorescence (SF)[9,10] could enable this kind of quantum control by producing ultrafast (few-picosecond), high-brightness, coherent photon bursts. However, to date, SF has only been observed under optical pumping and has never been demonstrated in a scintillator exposed to ionising radiation. The demonstration of SF under ionizing radiation would not only expand the fundamental understanding of collective light-matter interactions but also aligns with a growing interest in quantum-based radiation detection technologies for achieving enhanced control and information extraction from ionizing events[8,11]. In high-energy physics, such approaches could offer access to new observables and improved resolution of particle dynamics. More broadly, they may open up transformative opportunities in fields where speed, sensitivity, and miniaturization are key technological drivers.

A route beyond this structural bottleneck emerges from lead halide perovskite (LHP) nanocrystals (NCs)[12-14], particularly in the fully inorganic form $CsPbX_3$ (where X is either Cl, Br or I), that combine: *i)* scalable[15] and low-cost fabrication*, ii)* the high interaction cross-sections typical of high-Z inorganic scintillators[2,16,17]; and *iii)* exceptional defect tolerance and radiation hardness[18,19]. Individual LHP NCs exhibit ultrafast sub-nanosecond scintillation[20-24] driven by the recombination of multiexcitons formed under ionizing excitation[25-28] (as similarly observed in chalcogenide NCs[29-31]), which is further enhanced and accelerated under mild cryogenic conditions by the formation of giant transition dipoles (giant oscillator strength, or GOS) delocalized over the entire NC volume[32,33]. GOS causes a dramatic increase in the radiative recombination rate of both single and multiexcitons, while simultaneously boosting the efficiency of the emission process to its fully radiative limit (see sketch in **Figure 1a**). The result is scintillation radiative lifetimes as short as ~200 ps with near-unity yield from isolated NCs, a highly desirable feature for fast-timing applications[33].

A particularly powerful asset of LHP NCs for realising the quantum-scintillation regime is their propensity to self-assemble into highly ordered structures known as superlattices (SLs). A high degree of monodispersity and shape uniformity of LHP NCs, offered through the latest synthetic advancements[34-36], allowed the production of various single-component NC SLs[37-40] as well as multicomponent NC SLs containing different NC types[41-44]. Among semiconductor NCs, three-dimensional (3D) $CsPbBr_3$ NC SLs were the first to exhibit SF[37,39,40,45], a phenomenon previously observed in a limited number of non-colloidal systems[9,10]. NC SLs thus emerge as a desirable platform for harnessing SF for real-world applications, owing to the broad know-how in their lattice engineering, facile patterning, scalability and integration[46].

In NC-SLs, SF arises from the spontaneous phase-locking of individual resonant NC transition dipoles[37,40,47] (see **Figure 1a**), a phenomenon facilitated by the size selection occurring during the SL formation process[38,48,49]. This phase-locking leads to the emergence of a macroscopic correlated state with energy lower than the band-edge exciton of isolated NCs, whose oscillator strength scales with the number of coupled NCs[45,50], *N*. The decay of this correlated state, confined within a sub-wavelength volume, results in delayed photon bursts where both emission intensity and decay rates scale as $N^2$ and $N$, respectively[37,40,47,51]. In combination with the intra-NC GOS effect, the delayed formation and decay of these inter-NC coupled states in SLs lead to the emergence of spectrally narrower emission features with picosecond-fast decay times red-shifted by up to ~90 meV from the excitonic absorption of uncoupled NCs and without concurrent ground-state absorption, thus improving the potential for their use in high-precision detection systems.

This unique property of LHP SLs, particularly among inorganic colloidal nanostructures, has already been proposed for quantum information[52] and lasing[53,54] applications. Demonstrating such coherent emission under ionising excitation would mark a paradigm shift: the scintillator becomes an active, many-body quantum system whose response can be engineered rather than empirically accepted. Designed in this way, reabsorption-free, picosecond-fast nanoscintillators could potentially be integrated into more complex "metascintillator" architectures, combining bulk scintillator crystal layers with nanocrystal-based systems to enhance sensitivity and spectroscopic capabilities for high-energy detection[55].

Despite the potential and early-stage efforts[56], SF under ionizing radiation has not yet been experimentally demonstrated. Furthermore, arranging $CsPbBr_3$ NCs into two-dimensional (2D) SLs may significantly influence in-plane collective interactions by reducing the coupling dimensionality from 3D networks to 2D sheets, possibly leading to enhanced anisotropic coherent effects and further tuning of the scintillation response.

In this paper, we present the first experimental evidence of SF under ionizing radiation from 2D-like self-assembled superlattices of 7.6 nm $CsPbBr_3$ NCs. The SLs are characterized by a high degree of ordering and extended domain dimensions, as manifested by electron microscopy and electron diffraction studies. Photoluminescence (PL) measurements confirm superfluorescent behaviour in such SLs at cryogenic temperatures, characterized by a narrow, red-shifted emission peak with a radiative lifetime as short as 40 ps. This is accompanied by a significant increase in emission intensity that saturates below 30 K, consistent with previous optical studies[57,58]. X-ray-excited radioluminescence (RL) experiments on the same SL show that SF is the dominant emission pathway even under ionization conditions, marking the first instance of SF-driven scintillation. Importantly, optical spectroscopy experiments as a function of excitation fluence, together with time resolved scintillation measurements, reveal that ionizing excitation induces SF at high cryogenic temperatures (80-100 K) similar to what can be achieved with high-fluence optical excitation, with a scintillation lifetime as fast as ~40 ps. As a result, over 80% of the scintillation photons originate from SF due to the high exciton density (within the decay time) created by ionizing radiation, despite the restriction of the coherence domain due to dynamic thermal disorder, which makes the SF-driven scintillation process technologically accessible. At the same time, because of the build-up time required for the formation of the correlated SF state, the optical absorption spectrum of the SL is dominated by the uncoupled domains (absorbing at higher energy with respect to the SF), resulting in no spectral overlap between the SF emission and the absorption edge and consequent absence of reabsorption losses. To gain further insights into the interaction mechanisms between ionising radiation and the NCs within the SL, we complement our experimental observations with Monte Carlo simulations using the Geant4 toolkit[74,75,76]. These simulations were designed to replicate the geometry and composition of the SLs under investigation, allowing to estimate the spatial distribution of the energy deposited by incoming radiation. We thus modelled individual cubic SLs composed of $CsPbBr_3$ NCs, with nanometric resolution that reflects the actual arrangement and spacing of the NCs. By simulating both electrons and X-rays, we evaluated how energy is deposited across the lattice and how this energy distribution evolves depending on the radiation type and energy.

These results open new avenues for the development of ultrafast reabsorption-free scintillator metasolids, which hold potential for time-of-flight radiation detection technologies in both medical imaging and high-energy physics, where their ability to deliver fast, precise timing and high resolution can lead to significant advancements in radiation detection and imaging capabilities.

**Fabrication and structure of 2D $CsPbBr_3$ nanocrystals superlattices.**

A high degree of monodispersity and shape uniformity of SL building blocks is an essential prerequisite for the fabrication of long-range ordered SLs[59], enabling reliable experimental study. To achieve this, $CsPbBr_3$ NCs were synthesized by means of a facile, room-temperature approach assisted by trioctylphosphine oxide (TOPO) and diisooctylphosphinic acid (DOPA), which act as weakly coordinating ligands facilitating NC growth[36]. To stabilize the NCs in colloidal solution and promote their self-assembly, the labile TOPO and DOPA ligands were exchanged for stronger binding ligands. Specifically, the crude solution was first treated with a mixture of oleyl amine (OLAm) and oleic acid (OLA), followed by the addition of didodecyldimethylammonium bromide (DDAB) to improve the NCs' environmental stability[39,60,61]. The synthesis yielded highly monodisperse 7.6±0.7 nm $CsPbBr_3$ NCs with a sharp cubic shape, as evident from transmission electron microscopy (TEM) imaging (**Figure 1b**). The absorption spectrum of the NC colloidal solution exhibits multiple exciton transitions (**Figure 1c**), characteristic for the employed synthesis[36], whereas the narrow NC size-dispersion is reflected in a PL full width at half maximum (FWHM) of 16.7 nm.

As SF measurements require highly ordered SLs with extended domain sizes, we selected a facile drying-mediated approach on a tilted support as a method for SL fabrication[62] (**Figure 1d**). Unlike the deposition of 3D SLs, this method facilitates the directional alignment of SL domains on the substrate, *i.e.* NCs within different SL domains exhibit uniform preferential orientation. Besides, with this drying-mediated approach, the SL deposition is enabled on various substrates, including TEM grids or silicon nitride ($SiN_x$) membranes, using nonpolar solvents compatible with $CsPbBr_3$ NCs[41,44].

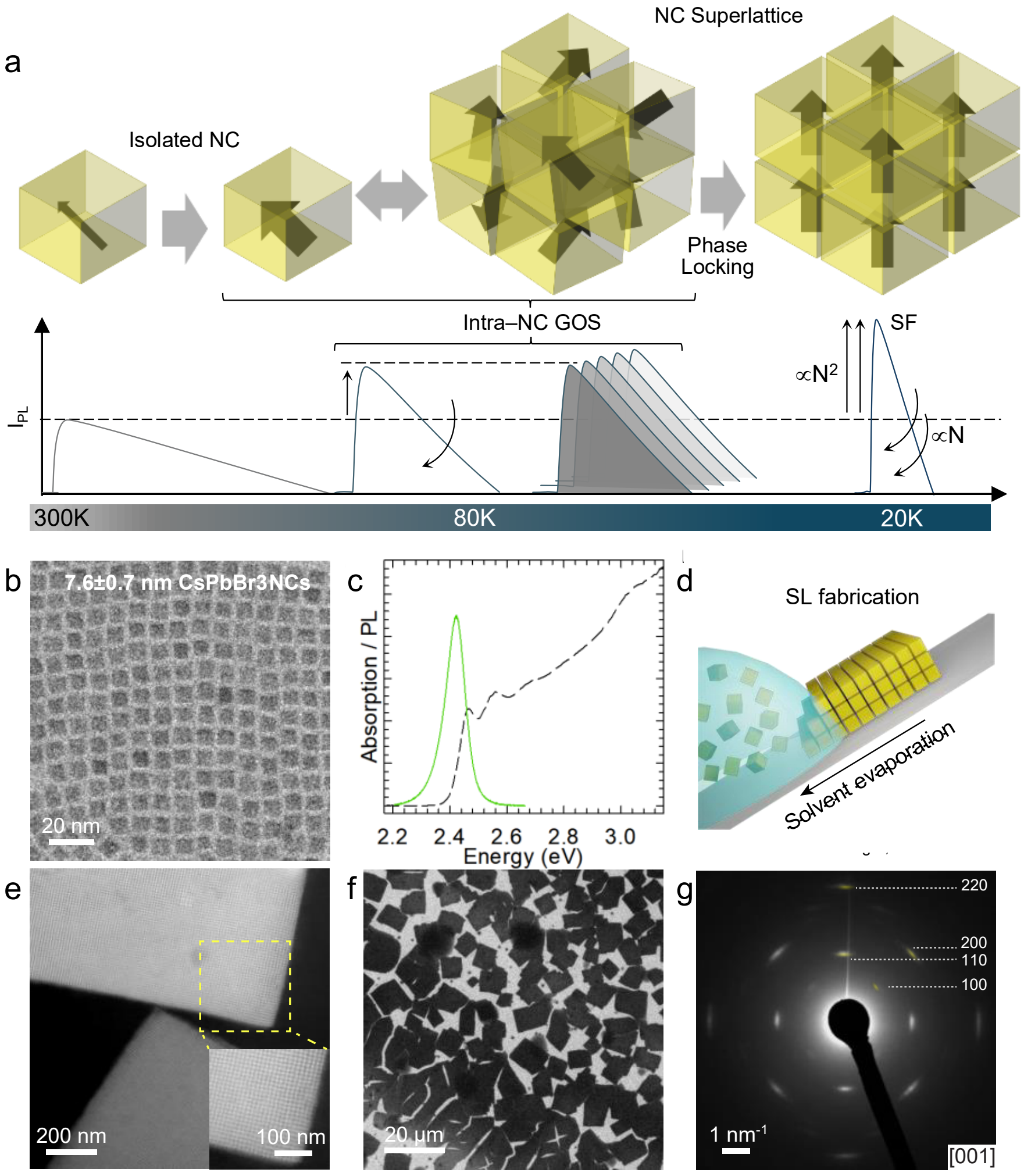

**Figure 1. $CsPbBr_3$ nanocrystal superlattices (NC SLs). a,** Sketch showing the progressive emergence of cooperative processes in isolated NCs and in NC SLs at cryogenic temperatures, leading to acceleration of the radiative dynamics and enhancement of the luminescence. The 'intra-NC' giant oscillator strength (GOS) effects, both in isolated NCs and in assembled NC superlattices, build up below room temperature resulting in faster single emitting dipoles with near unity quantum yield. At lower temperatures, quantum phase-locking of excited states within the coherence volume of a NC SL produces a SF photon burst whose intensity and decay rate scale proportionally to the number of coupled dipoles, *N*. **b,** TEM image of starting cubic $CsPbBr_3$ NCs with an edge length of 7.6±0.7 nm. **c**, Optical absorption (black line) and PL spectra (green line) of the $CsPbBr_3$ NC colloidal solution in toluene. **d**, Schematic illustration of SL fabrication by means of a drying-mediated approach, whereby the NC colloid is slowly evaporated over a tilted support, *e.g.* a TEM grid or $SiN_x$ membrane. **e**, Low-magnification dark-field STEM image of two neighbouring SL domains with (inset) dark-field STEM image of the area outlined with a dashed yellow line, demonstrating highly ordered NCs in simple cubic (*sc*) packing. **f**, Low-magnification TEM image of $CsPbBr_3$ NC SLs deposited on $SiN_x$ membrane, displaying extended SL domains with an average area of *ca.* 100 μm$^2$. **g,** Wide-angle electron diffraction (WAED) pattern of a $[001]_{SL}$-oriented SL domain, with reflections corresponding to preferential orientation marked with yellow. The sharp reflections highlight the orientation uniformity of SL domains.

Owing to the NCs' sharp cubic shape, self-assembly of 7.6 nm $CsPbBr_3$ NCs resulted in the formation of 2D SLs with simple cubic (*sc*) packing[63], as confirmed by dark-field scanning transmission electron microscopy (DF-STEM) imaging (**Figure 1e**), consistent with previous reports on similarly sized $CsPbBr_3$ NC assemblies[37,39]. The resulting extended 2D-like SLs exhibited distinct domains with a high degree of positional and orientational NC ordering across the substrate. On transparent $SiN_x$ membranes, individual SL domains extended over areas exceeding 100 $\mu m^2$, with a substrate coverage of ~60% (**Figure 1f**). In between the SL domains, few NCs were observed, *i.e.* the majority of NCs was located within the SL structure. The wide-angle electron diffraction (WAED) pattern displays intense, sharp sets of 100, 110, 200, and 220 reflections, indicating preferential orientation of NCs along the $[001]_{SL}$ direction (**Figure 1g**, subscript 'SL' denotes sets of equivalent SL directions).

**Superfluorescence of 2D $CsPbBr_3$ nanocrystal superlattices under optical excitation.**

Prior to the scintillation measurements, we performed time- and spectrally-resolved PL measurements as a function of temperature to study the emission mechanism of this new class of SLs and to verify that they exhibit SF similar to those produced by other methods reported in the literature[37,58,64,65]. Specifically, as previously observed, the characteristic signatures of SF in $CsPbBr_3$ NC SL are the delayed emergence of a low-energy emission due to coupled NCs (~60 meV below the PL from uncoupled NCs) with lifetime gradually decreasing with decreasing temperature (down to ~40 ps) and fluence scaling of SF intensity ($I_{SF}$) approaching the quadratic theoretical limit ($I_{SF}$~$f^{1.5}$ in ref [37]). **Figure 2a** shows the contour plots of PL(E,t) collected by a streak camera (resolution <10 ps) at $T$=180 K, 70 K, and 5 K under low optical excitation such that the average exciton population was <0.1 exciton per NC. For comparison, **Figure 2b** shows the analogous contour plots of a disordered film of the same NCs dispersed in a polystyrene matrix. The complete set of spectra and their decay curves within the investigated temperature range is shown in **Figure 2c** and **2d**, and **Supplementary Figure 1**. The respective parameters of integrated PL intensity ($I_{PL}$), lifetime and linewidth (*FWHM*) are quantified in **Figure 2g-i**. **Figures 2e** and **2f** show a detail of the time rise of SF and PL intensity at 5 K from uncoupled NCs in the SL and the corresponding emission spectra at increasing delay times (0-24 ps) extracted from **Figure 1a**. At 180 K, both samples exhibit a single PL peak at 2.43 eV, characteristic of isolated $CsPbBr_3$ NCs, which intensifies and red-shifts with decreasing temperature as a result of bandgap renormalization[66] and suppression of non-radiative decay channels. At the same time, the film of isolated NCs shows the shortening of the luminescence lifetime due to the gradual build-up of the oscillator strength of individual NCs, *i.e.*, the appearance of GOS, approaching saturation at about 70 K, in agreement with recent reports[33,67]. The emissive component from isolated NCs of the SL follows the same trend, but with a basically constant lifetime at about 100 ps (not shown), due to the dominant effect of dynamic quenching by the copper substrate of the TEM grid on which it is placed. More importantly, at about 90 K, the SL shows the onset of a second, narrower emission peak at about 2.33 eV, shifted by ~60 meV from the corresponding emission of isolated NCs at 2.39 eV. This narrow emission is consistent with the SF reported for similar SLs[37] and is due to collective emission from coupled NCs in higher-order local domains. Consistently, reducing the thermal disorder counteracting phase coupling by further lowering the temperature causes its strong enhancement to become the largely dominant emission, accelerating to about 40 ps lifetime and narrowing (**Figure 2g-i**). In contrast, lowering the temperature below 80 K has little effect on the intensity and dynamics of the decoupled emission of both SL and polystyrene NC film, consistent with the saturation of the oscillator strength and emission yield.

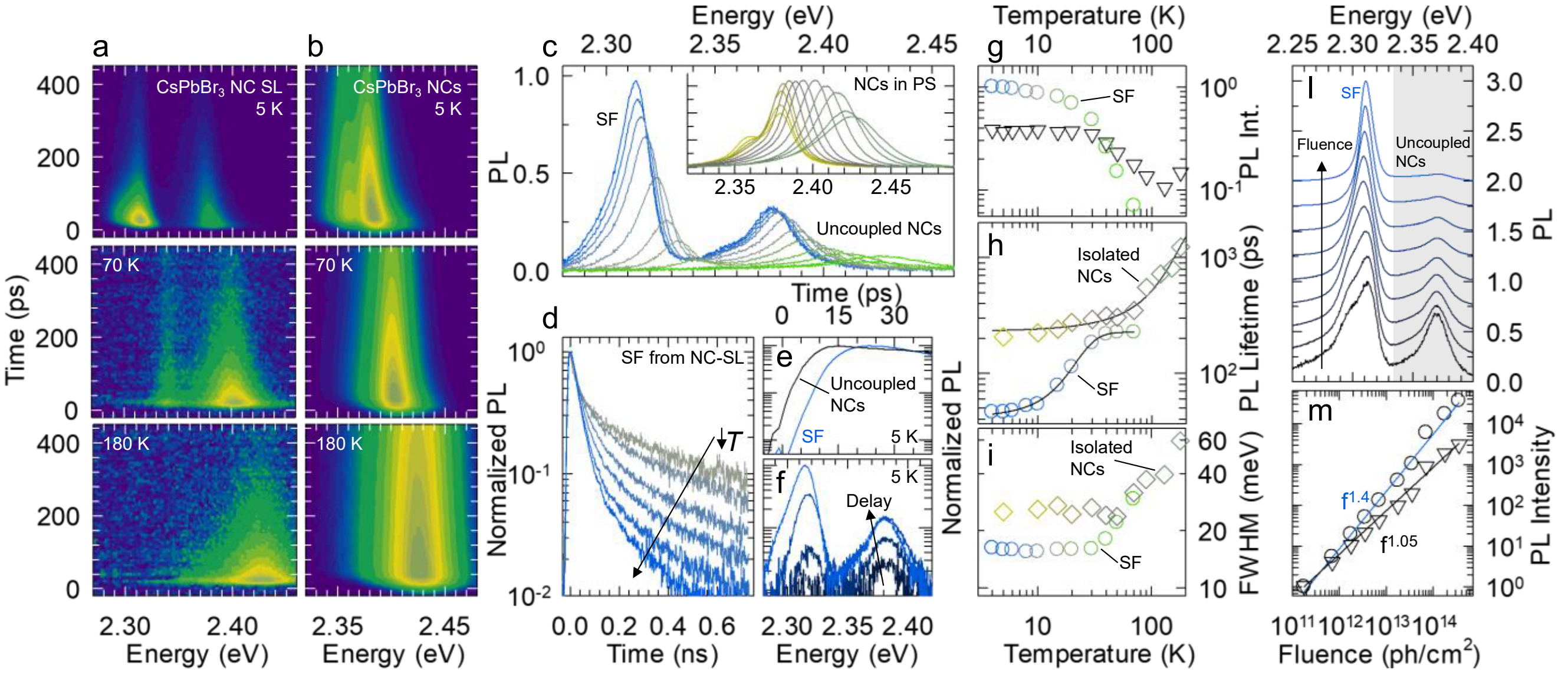

**Figure 2. Optically excited SF from $CsPbBr_3$ NC superlattices.** Contour plots of spectrally and time-resolved luminescence at controlled temperature (5, 70, and 180 K from top to bottom) of (**a**) $CsPbBr_3$ SL and (**b**) isolated $CsPbBr_3$ NCs dispersed in polystyrene. **c,** PL spectra of $CsPbBr_3$ NC SL measured in the 5-180 K temperature range. Inset: PL spectra of isolated NCs in polystyrene. The spectra have been normalized at the corresponding maximum at 5 K. **d,** Normalized PL decay traces of the superfluorescence (SF) component from $CsPbBr_3$ NC SL in the 5-40 K temperature range. **e,** PL rise signals of the SF (blue) and uncoupled (black) components extracted from top panel 'a'. **f,** PL spectra extracted from the top panel 'a' at progressively longer delay times (0, 3, 10, 24 ps). **g,** Intensity of the SF component (circles) from NC SL and corresponding PL from isolated NCs (triangles) as function of temperature. **h,** PL lifetimes of SF component (circles) and of isolated NCs in polystyrene (diamonds) at controlled temperature. The lines are a guide for the eye. **i,** Emission spectral width – expressed as FWHM – of the SF component in 'c' (circles) and of the PL for isolated NCs in polystyrene (diamonds). **l,** PL spectra of $CsPbBr_3$ NC SL at 5 K at increasing excitation fluence. The spectra have been normalized to their maxima and vertically shifted for clarity. **m,** Integrated PL intensity from panel 'l' for the SF (circles) and uncoupled NC PL (triangles) components as a function of excitation photon fluence. The same temperature colour scheme applies throughout the figure.

Also consistent with previous results[37], the SF intensity shows a delay time of ~6 ps from the corresponding uncoupled PL due to the formation of the correlated SF state, as highlighted by the respective decay traces and spectra collected at increasing times after the excitation pulse shown in **Figure 2e** and **2f**. As will be shown later in this work (*vide infra*), the fact that the superfluorescent state does not exist in the ground state but is formed gradually over time is advantageous for the realization of superfluorescent scintillators because it separates the optical absorption due to NCs before their coupling from the SF due to coupled NCs. Taken together, these spectroscopic signatures confirm the collective emission in SL of $CsPbBr_3$ NCs and the gradual increase in the number of coupled NCs at progressively lower temperatures. Further confirming this attribution, measurements of PL at 5 K as a function of excitation fluence, $f$ (**Figure 2l** and **2m**), show a linear trend ($I_{PL}$~$f^{1.05}$) for the emission from isolated NCs and proportional to $f^{1.4}$ for the SF component, for which a theoretical quadratic limit trend in the absence of non-radiative losses is predicted in agreement with ref.[37]

**Superfluorescent scintillation from of 2D $CsPbBr_3$ nanocrystal superlattices.**

After confirming the SF process and validating this class of SLs as a model system for the study of SF scintillation, we performed temperature-dependent RL measurements in the 20-180 K range using a tungsten cathode X-ray tube as the excitation source, which produces a continuous bremsstrahlung spectrum with a mean energy of 9.6 keV. The collected spectra are shown in **Figure 3a**. At 20 K, the RL spectrum exhibits two distinct components, which spectrally match the band-edge emission of isolated NCs and the coupled SF state observed above, with the latter accounting for more than 92% of the total RL signal. This confirms that, as in previously studied disordered NC films[28], the interaction with ionizing radiation in $CsPbBr_3$ SLs is intrinsically equivalent to high-fluence optical excitation, implying that, in RL measurements, the energy deposited by X-ray photons alone is sufficient to trigger and sustain collective phenomena, making SF the dominant scintillation mechanism. In this regard it is important to note that ionizing interaction generates multiple excitons in individual NCs following a carrier multiplication process[25,29], which is particularly advantageous for the SF regime because the collective oscillator strength of coupled NCs increases with the exciton density within the coherence domain. As shown in **Supplementary Figure 2**, the scintillation of isolated NCs shows an average excitonic population $\langle N \rangle \sim 1.5$ in agreement with previous reports[25]. Consistently, the RL shows the same spectral evolution (see markers in **Figure 3a**), as well as similar intensity trends and relative intensity contribution between the SF and isolated NC components, as the temperature-dependent PL spectra collected under high fluence excitation (~80 μJ/cm², *i.e.* $\sim 1.8 \times 10^{14}$ ph/cm², **Figure 3b**), selected to match the $\langle N \rangle$ produced under X-rays. The corresponding integrated PL and RL intensities, along with the branching ratio between the coupled and uncoupled emission, are quantified in **Figures 3c** and **3d,** respectively (the complete set of PL spectra as a function of fluence is reported in **Supplementary Figure 3**). These figures highlight a particularly relevant aspect of the parallelism between high irradiance optical and ionizing excitation, namely that both lead to essentially instantaneous formation of high excitation densities within the coherence domain of SL. As a result, SF scintillation under X-rays (as well as its optical counterpart) is observed even at high temperatures despite increasing dynamical disorder. Specifically, the SF component consistently accounts for more than 70% of the total RL signal in the experimental temperature range, and shows promising performance even at liquid nitrogen temperatures, a mild cryogenic regime that is already technologically accessible. For example, at ~80 K, the relative contribution of the SF component has already approached its maximum of ~90%, while the total RL intensity is only reduced by a factor of two compared to 20 K. A crucial aspect of the SF regime is the radiative acceleration of emission dynamics, which improves scintillation timing. To measure scintillation kinetics below ~100 ps, as expected based on PL dynamics in **Figures 2d** and **2h**, we performed two-photon cross-correlation measurements[68] under continuous-wave X-ray excitation (see sketch in **Figure 3e**). This configuration is necessary for time-resolved RL because the pulse width of pulsed

X-ray tubes typically exceeds 120 ps[69]. **Figure 3f** reports the measured $g^{(2)}$ curve at 20 K showing a positive peak that decreases to zero at increasing delay times, as expected from photon bunches produced by SF. From analysing the $g^{(2)}$ kinetics, we obtained a scintillation lifetime of ~40 ps (see Methods section for details) confirming the potential of SF scintillation for fast timing applications. To the best of our knowledge, this is the fastest scintillation time ever reported under ionizing radiation. Only Cherenkov-based processes are intrinsically faster[70,71], yet their extremely low and highly fluctuating photon yields (typically <20 photons per γ-event in PET scanners for $Bi_4Ge_3O_{12}$ (BGO) crystals) limit their practical use as time taggers to enhancing the timing performance of the scintillating host material in which they are generated, provided that advanced pulse-shaping and filtering algorithms are employed[72]. Finally, the close match between the scintillation and high-fluence optically excited PL enables us to study the temperature- and fluence-dependence of SF in greater detail using the time-correlated single-photon counting technique. **Supplementary Figure 4** shows the decay curves at 50 K as a function of femtosecond laser excitation fluence, as well as the decay curves at $\langle N \rangle$ ~1.5 as a function of temperature, together with the corresponding effective lifetimes. The complete set of PL decays and the details analysis of the SF dynamics, including the fit convoluted with the detector's response time, are presented in **Supplementary Figure 5**. Consistent with the expected SF trend[37-40], decreasing the temperature and increasing the fluence results in accelerated decay kinetics compared to those observed at low fluence in **Figure 2**. This is due to the simultaneous increase in the average excitonic population per NC (i.e. activation of the multiexcitonic regime) and increased number of excited NCs within the coherence domain.

Overall, the spectral and timing behavior provides the first evidence that $CsPbBr_3$ SLs can operate as cryogenic scintillators in which the RL originates predominantly from a coupled SF state, resulting in emission that is further red-shifted by ~60 meV from isolated NCs and radiatively accelerated with lifetimes that can reach a few tens of ps depending on the number of coupled emitters. These are the key requirements for a dense, ultrafast scintillator capable of emitting intense photon bursts spectrally decoupled from the host (emitter) matrix. Stopping high-energy photons, such as those used in medical imaging, requires scintillators a few centimeters thick with negligible optical losses due to self-absorption to maximize light output to the detector. To further support this application, **Figure 3g** shows the temperature evolution of the optical absorption profile of $CsPbBr_3$ SLs deposited on an ultra-thin transparent $SiN_x$ membrane. At room temperature, the high monodispersity of the NCs allows to resolve the first two excitonic transitions at 2.46 eV and 2.56 eV, in excellent agreement with previous reports on highly monodisperse colloidal solutions of NCs with similar size[36,73]. As temperature decreases, both transitions shift to lower energy, following thermal expansion of the NCs lattice, and increase in intensity due to the onset of intra-NC GOS. At 20 K, the lowest 1S transition reaches ~2.41 eV. Notably, the corresponding RL spectrum shows a residual RL from uncoupled NCs at 2.37 eV, consistent with the typical Stokes shift of $CsPbBr_3$ NCs[74], and a dominant SF peak at 2.31 eV, setting the overall spectral shift between the SF scintillation and the SL absorption edge up to 100 meV, which essentially completely hinders reabsorption.

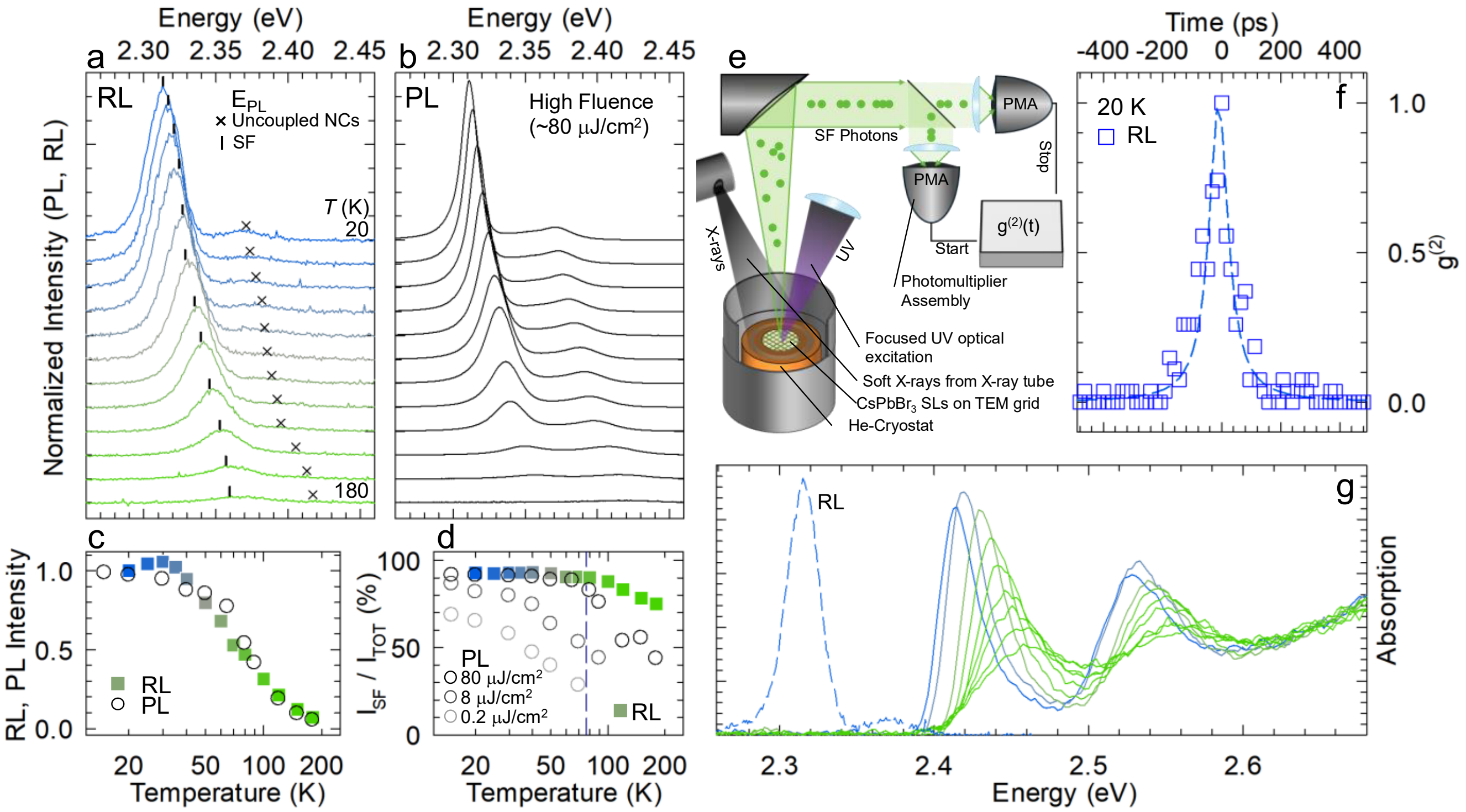

**Figure 3. SF under ionizing excitation. a,** Radioluminescence (RL) spectra of $CsPbBr_3$ SLs at controlled temperatures from 180 to 20 K (from bottom to top). The spectra have been normalized to the spectrum maximum at 20 K and vertically shifted for clarity. The vertical ticks and crosses correspond to the spectral positions ($E_{PL}$) of SF and isolated emissions, respectively, from the corresponding PL spectra under high fluence excitation, as shown in '**b**'. **c,** Normalized integrated PL (circles) and RL (squares) intensity as a function of temperature as extracted from panels 'a' and 'b'. **d,** Relative intensity of the SF component (expressed as $I_{SF}/I_{total}$) as a function of temperature from RL (squares) and PL (circles) measurements at increasing fluences (~0.2, 8, and 80 µJ/cm²). The vertical dashed line corresponds to the technologically relevant liquid nitrogen temperature. **e,** Schematic illustration of the cross-correlation setup in the Hanbury Brown and Twiss configuration, together with a sketch of sample's excitation (X-rays or UV light) and SF light collection within a He-cryostat. **f,** Second-order RL (squares) correlation function measured for $CsPbBr_3$ NC SLs at 20 K along with the best fit with a Lorenzian function (blue line) featuring a FWHM of 80 ps. **g)** A set of representative optical absorption spectra at decreasing temperatures (from 290 K to 20 K, from green to blue) for $CsPbBr_3$ SLs. The corresponding RL spectrum at 20 K is shown as a dashed line.

To gain deeper insight into the mechanisms underlying energy deposition in NC SLs, we performed Monte Carlo simulations using the Geant4 toolkit, starting from a representative SL geometry composed of 7.6 nm $CsPbBr_3$ NCs separated by 1.4 nm OLAm ligands. **Figure 4a** illustrates the case of a cubic SL with ~100 nm edge length (containing 1728 NCs), irradiated by a uniform flux of X-rays matching the spectral distribution of the source used in **Figure 3**. The projection of the energy deposited by external X-rays in the XY plane is shown in **Supplementary Figure 6**. These simulations reveal how incident radiation interacts with the SL's architecture and provide a spatially resolved map of energy deposition within the lattice. Notably, energy is absorbed predominantly in the NCs rather than the surrounding OLAm, indicating the dominant role of the perovskite domains in radiation-matter interaction.

To further investigate how energy spreads following a primary interaction, we simulated the generation of energetic electrons at the centre of the cubic SL and tracked their energy-loss trajectories (**Figure 4b**). While this setup is not fully representative of physical X-ray interaction - since X-rays do not typically generate electrons at the SL centre - it offers valuable insight into local energy propagation. Unlike diluted solutions or polymeric NC dispersions, where most of the energy dissipates into the inert matrix[75], the densely packed SL architecture promotes localized energy retention, enhancing local exciton generation and radiative recombination. Although these detailed simulations are informative, they become computationally expensive as the number of NCs increases. For realistic SLs containing ~20 million NCs, explicitly modelling each particle becomes unpractical. A feasible solution is to adopt a simplified approach in which the SL is approximated as a homogeneous medium with equivalent mass fractions of $CsPbBr_3$ and OLAm. We validated this method by comparing the energy deposited in the mixed-volume model with that of explicit NC simulations (**Supplementary Figure 7**), finding that the difference becomes negligible for NC counts above $\sim 10^6$. Based on this validation, we simulated a flat SL of ~100 $\mu m^2$ surface area and 141 nm thickness, corresponding to a realistic lattice of ~20 million NCs. The resulting mean deposited energy per interacting X-ray photon was 3.97 keV. Combined with the quantitative RL measurements shown in **Figure 3a**, this enables us to estimate the LY of our SLs under X-ray excitation, yielding LY = 64,000 ± 20,000 photons/MeV at 20 K (see Methods for details). This value is higher than that of standard $CsPbBr_3$ NC films (~10,000-15,000 photons/MeV) and is consistent with a scintillation process that is largely radiative. Nevertheless, the theoretical maximum LY for $CsPbBr_3$ remains ~160,000 photons/MeV, indicating that photocharge-to-exciton conversion in the SL is still incomplete[76,77]. It is important to note that the LY estimation should be interpreted with caution, as accurate experimental measurement of the deposited energy is currently not feasible. Nevertheless, the RL system was calibrated assuming Lambertian emission from the sample, which represents a conservative (lower-bound) assumption for the collection geometry compared to fully isotropic emission.

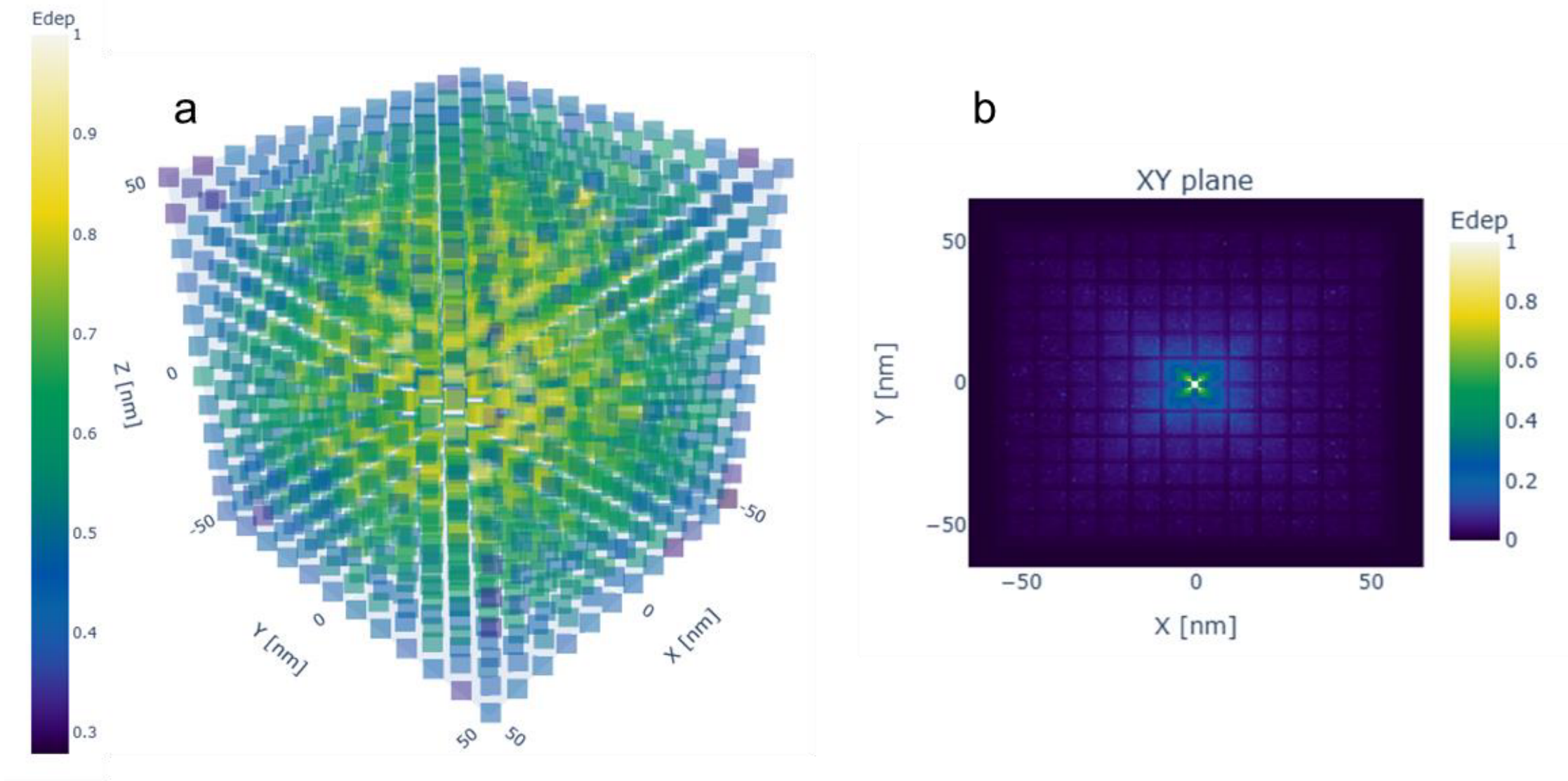

**Figure 4. Monte Carlo simulation of energy deposition in nanocrystal SLs. a,** Three-dimensional rendering of a simulated cubic SL composed of 1728 $CsPbBr_3$ NCs regularly arranged and embedded in a surrounding OLAm matrix. Each NC is colour-coded based on the energy deposited (normalized over the maximum released) per incident X-ray (excitation source outside the SL), illustrating the spatial distribution of the energy within the lattice. **b,** Orthogonal projections onto the plane XY of the energy deposited (normalized over the maximum released) for the case of internally generated photoelectrons in the centre of the same cubic SL represented in 'a'.

## Conclusions

In conclusion, we successfully fabricated macroscopically two-dimensional SLs from highly monodisperse 7.6 nm $CsPbBr_3$ NCs, stabilized with OLAm/OLA/DDAB ligands that promote controlled self-assembly. Transmission electron microscopy and electron diffraction analyses revealed highly ordered, orientationally locked NCs arranged in a cubic lattice. Remarkably, the resulting SLs exhibited large, continuous domains (~100 μm$^2$) with high areal coverage on TEM grids and $SiN_x$ membranes. We employed these $CsPbBr_3$ SLs as a model platform to investigate collective optical phenomena under ionizing radiation. We demonstrated that under X-ray excitation, the RL of the SLs is dominated by SF, accounting for over 90% of the total emission with lifetimes as fast as ~40 ps at 20 K. This ultrafast response persisted with minimal degradation up to and beyond 80 K. These findings indicate that ionizing excitation can drive the system into a collective emission regime, establishing SF as the primary scintillation mechanism in this class of hybrid metamaterials. Complementary temperature-dependent optical absorption measurements revealed clear signatures of the giant oscillator strength effect at the band edge and confirmed the absence of spectral overlap between the SF emission and the absorption edge, indicating negligible self-absorption losses. To understand how energy is deposited in the SL under high-energy irradiation, we performed Monte Carlo simulations using Geant4, modelling the NC lattice embedded in an OLAm matrix. Simulations showed that the majority of the total energy deposited by secondary interactions is found within closed-packed NCs along thin tracks consistent with the spatial scales of collective coupling. Taken together, our results establish $CsPbBr_3$ SLs as a new generation of solution-processable scintillators, in which collective radiative phenomena govern the response under both optical and ionizing excitation. This paves the way for ultrafast detection architectures based on coherently emitting NC assemblies with minimal reabsorption losses and fundamentally new mechanisms for signal generation and processing.

## Methods

*Synthesis of CsPbBr3 nanocrystals and SL growth:*

$CsPbBr_3$ NCs were fabricated employing the adopted procedure from Akkerman et al.[36]. $PbBr_2$-TOPO stock solution (0.04M) was prepared by dissolving $PbBr_2$ (1 mmol, 367 mg, Sigma-Aldrich) and trioctylphosphine oxide (TOPO, 5 mmol, 1.933 g, Strem) into octane (5 mL, Roth) at 110 $^{o}$C in a 40 mL vial. After all $PbBr_2$ dissolved, the solution was allowed to cool down to room temperature and hexane (20 mL) was added. The stock solution was then filtered with 0.2 μm PTFE filter.

Cs-DOPA stock solution (0.02M) was prepared by dissolving $Cs_2CO_3$ (0.3 mmol, 100 mg, Sigma-Aldrich) and diisooctylphosphinic acid (DOPA, 3.15 mmol, 1 mL, Sigma-Aldrich) into octane (2 mL, Roth) at 110 $^{o}$C in a 40 mL vial. After all $Cs_2CO_3$ dissolved, the solution was allowed to cool down to room temperature and hexane (27 mL) was added. The stock solution was then filtered with 0.2 μm PTFE filter.

Oleyl amine (OLAm, Strem, distilled) and oleic acid (OLA, Sigma-Aldrich) ligand solution was prepared by mixing OLAm (0.33 mL, 1 mmol) and OLA (0.316 mL, 1 mmol) in anhydrous hexane (2.5 mL).

Didodecyldimethylammonium bromide (DDAB) solution (0.215 M) was prepared by dissolving DDAB (Sigma-Aldrich, 0.3 g, 0.65 mmol) in anhydrous toluene (3 mL, Sigma-Aldrich).

*Synthesis of 7.6 nm $CsPbBr_3$ NCs.* In a 25 mL round-bottom flask, $PbBr_2$-TOPO stock solution (1.5 mL) was combined with hexane (1 mL) upon vigorous stirring (1400 rpm). Then Cs-DOPA stock solution (0.75 mL) was injected, and after stirring for 90s, OLAm and OLA ligand solution (1 mL) was added. The solution was allowed to stir for 60 s, and DDAB (30 μL) was injected. NC colloid was then concentrated with a rotary evaporator and washed with anhydrous ethyl acetate (EtOAc, 1:1 v/v). NCs were redispersed in anhydrous toluene and stored inside the glovebox.

*Fabrication of $CsPbBr_3$ NC SLs.* $CsPbBr_3$ NC SLs were prepared employing drying-mediated approach (Figure 1d). TEM grids (F/C-coated, Ted Pella, with the Formvar layer removed by immersing the grid in toluene for 10 s) or $SiN_x$ membranes (Agar Scientific, Norcada) were employed as the substrates. The NC colloid (5-10 μL, 2 μM) was placed along with 15 μL anhydrous toluene in a 2 mL vial with a solid support inside. The vial was then tilted in a vacuum chamber (pressure ~0.5 bar, room temperature), where the solvent slowly evaporated overnight.

*Microscopy characterization:*

TEM and STEM images, as well as WAED patterns, were collected with a JEOL JEM 2200FS electron microscope operating at an accelerating voltage of 200 kV. Image analysis was performed using ImageJ.

*Optical spectroscopy:*

The optical absorption spectra, also measured as a function of temperature, were collected using a Lambda 950 spectrophotometer (Perkin Elmer) equipped with an integrating sphere and coupled with a closed-circuit He cryostat. For low fluence ultrafast time-resolved PL (trPL) measurements the samples excited with frequency-doubled Ti:sapphire laser ($E_{exc}$ = 3.06 eV, pulse duration ~150 fs, repetition rate ~ 76 MHz) collecting the emitted light with a Hamamatsu streak camera (time resolution < 10 ps). For fluence-dependent PL measurements – both *cw* and time-resolved – the samples were excited with an amplified laser from Ultrafast Systems operated at 15 kHz, producing ~260 fs pulses at 3.06 eV when coupled an independently tuneable optical parametric amplifier from the same supplier. The emitted light was collected with a Horiba Triax 190 monochromator (grating 1200 lines/mm) coupled with a cooled Horiba Sincerity CCD for steady state PL, and a Cornerston 260 1/4 m VIS-NIR monochromator coupled with a PicoQuant PMA Hybrid Series-07 and a PicoHarp 300 time-correlated single photon counting unit. All the PL measurements were performed inside a closed-circuit cryostat operated in evaporated He atmosphere at ~10 mbar.

*Intensity cross-correlation measurement:*

Two-photon cross-correlation measurement was conducted using a standard Hanbury Brown and Twiss configuration with two matched PMA Hybrid single photon detector. The intensity correlation function ($g^{(2)}$) was generated using the two detectors as start and stop triggers for a PicoHarp 300 time-tagger. Negative correlation times were probed thanks to an electronic delay set for one of the two channels. The correlator's resolution is estimated to be ~70 ps – full width at half maximum (FWHM) – assuming a ~50 ps spread time for each detector. The SF dynamics have been calculated from the fitted FWHM according to:

$$\tau_{SF} \approx \sqrt{FWHM_{Meas}^2 - FWHM_{Instr}^2}$$

*Radioluminescence spectroscopy:*

The cryogenic radioluminescence measurements were performed with a closed cycle He cryostat exciting the sample with unfiltered X-rays produced by a Philips PW2274 X-ray tube with a tungsten target, equipped with a beryllium window, and operated at 20 kV to produce a continuous X-ray spectrum through bremsstrahlung effect. The scintillation light spectra were collected using a liquid-nitrogen-cooled, back-illuminated, UV-enhanced CCD detector (Jobin Yvon Symphony II), coupled to a monochromator (Jobin Yvon Triax 180) with a 600 lines/mm grating. The time response of scintillation light was measured using a pulsed X-ray source composed by a 405 nm ps-pulsed laser (Edinburgh Instruments, EPL-405) hitting the photocathode of an X-ray tube from Hamamatsu (model N5084) set at 40 kV. The scintillation was collected using a MicroHR spectrometer from Horiba (equipped with a 150 lines/mm grating) coupled with a PicoQuant PMA Hybrid Series-07 and a PicoHarp 300 for time-correlated single photon counting (time resolution ~120 ps).

*Light Yield Estimation:*

The light yield (LY) is defined as the ratio between the number of emitted photons and the energy deposited in the scintillator, typically expressed in MeV.

$$\mathrm{LY} = \frac{N_{\mathrm{ph}}}{E_{\mathrm{deposited}}}\ [\mathrm{photons\ MeV^{-1}}]$$

To convert the CCD counts from the radioluminescence (RL) spectrum into the photon number $N_{ph}$, we first measured the optical detection efficiency of the setup at the SF wavelength. A continuous-wave 532 nm laser was coupled into a multimode optical fiber with a core diameter of 600 µm, chosen to match the lateral dimensions of the sample, and with a numerical aperture equal to that of the first collection optic. The power emitted from the fiber tip was measured using a calibrated power meter and converted into the number of emitted photons per second. The detection efficiency η is then calculated as the ratio between the CCD counts (when measuring the light intensity from the fiber tip under the same grating, slit, and geometrical conditions used for the RL measurement) and the photon flux at the fiber tip. The sample was modelled as a Lambertian emitter to estimate the total number of photons produced per second.

To determine the energy deposited in the scintillator, we first calculate the X-ray tube photon flux $N_X$ [ph·cm$^{-2}$·s$^{-1}$] as:

$$N_X = \frac{D_{\mathrm{air}}}{\int \tilde{\Phi}(E)\left(\frac{\mu_{\mathrm{en}}}{\rho}(E)\right)_{Air} E\, dE}\text{, where } \int \tilde{\Phi}(\mathrm{E})\,\mathrm{dE} = 1$$

known the dose rate in air $D_{Air}$ [J kg$^{-1}$s$^{-1}$] (measured with a calibrated ionization chamber), the X-rays photon fluence spectrum $\tilde{\Phi}(E)$ (simulated using the XMI-MSIM package[78]), and the mass energy-absorption coefficients ($\frac{\mu_{en}}{\rho}(E)$, cm$^2$/g) of dry air (as extracted from NIST X-COM database[79]).

The number of incident X-ray photons on the sample is obtained multiplying $N_X$ for the area of the sample. The total energy deposited within the sample is obtained from the Geant4 simulation for that number of incident X-photons.

## Acknowledgments

This work was funded by Horizon Europe EIC Pathfinder program through project 101098649 – UNICORN, by the European Union -Next Generation EU, Mission 4 Component 1 CUP H53D23004670006 and CUP H53D23004500006, and through the Italian Ministry of University and Research under PNRR—M4C2-I1.3 Project PE_00000019 "HEAL ITALIA". This research is funded and supervised by the Italian Space Agency (Agenzia Spaziale Italiana, ASI) in the framework of the Research Day "Giornate della Ricerca Spaziale" initiative through the contract ASI N. 2023-4-U.0t.

**Author contributions**

S.B. and M.V.K. conceived this work. T.S. synthesized the NCs and assembled the superlattices under the supervision of M.B. and M.V.K..M.L.Z, A.F., F.B. and F.C. conducted the spectroscopic and scintillation experiments under the supervision of S.B. and F.M.. E.M. conducted the GEANT4 simulations under the supervision of L.G.. S.B., M.L.Z. and F.M. wrote the paper in consultation with all authors.

**Competing Interest**

The authors declare no competing interests.

# Supplemental Information

## Radiation-Triggered Superfluorescent Scintillation in Quantum-Ordered Perovskite Nanocrystal Superlattices

Matteo L. Zaffalon[1,2], Andrea Fratelli[1,3], Taras Sekh[4], Emanuele Mazzola[2,5], Francesco Carulli[1], Francesco Bruni[12], Maryna Bodnarchuk[4], Francesco Meinardi[1], Luca Gironi[2,5], Maksym V. Kovalenko[4]*, Sergio Brovelli[1,2]*

[1]*Dipartimento di Scienza dei Materiali, Università degli Studi di Milano Bicocca, via R. Cozzi 55, Milano, Italy.*
[2]*INFN - Sezione di Milano - Bicocca, Milano, Italy.*
[3]*Nanochemistry, Istituto Italiano di Tecnologia, via Morego 30, Genova, Italy.*
[4]*Department of Chemistry and Applied Bioscience, ETH Zürich, Zürich, Switzerland.*
*Laboratory for Thin Films and Photovoltaics and Laboratory for Transport at Nanoscale Interfaces, Empa – Swiss Federal Laboratories for Materials Science and Technology, Dübendorf, Switzerland.*
[5]*Dipartimento di Fisica, Università degli Studi di Milano-Bicocca, Piazza della Scienza, Milano, Italy.*

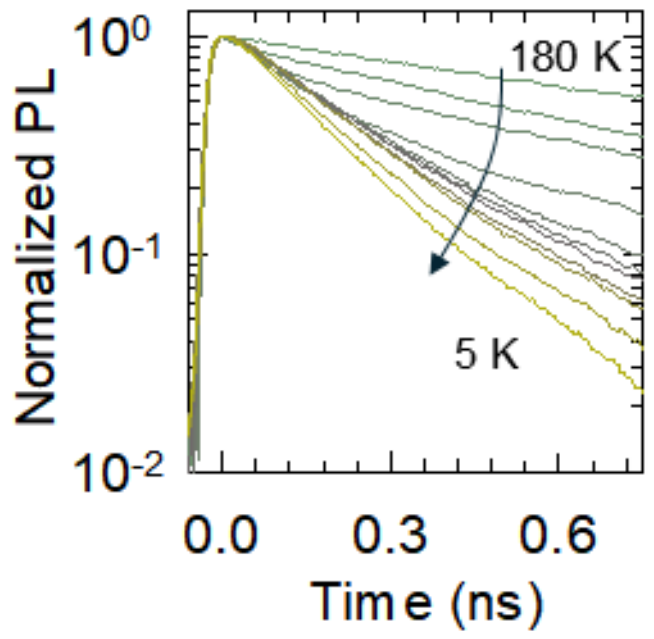

**Supplementary Figure 1:** Normalized PL decay traces of from isolated $CsPbBr_3$ NCs dispersed in a poly styrene matrix in the 5-180 K temperature range.

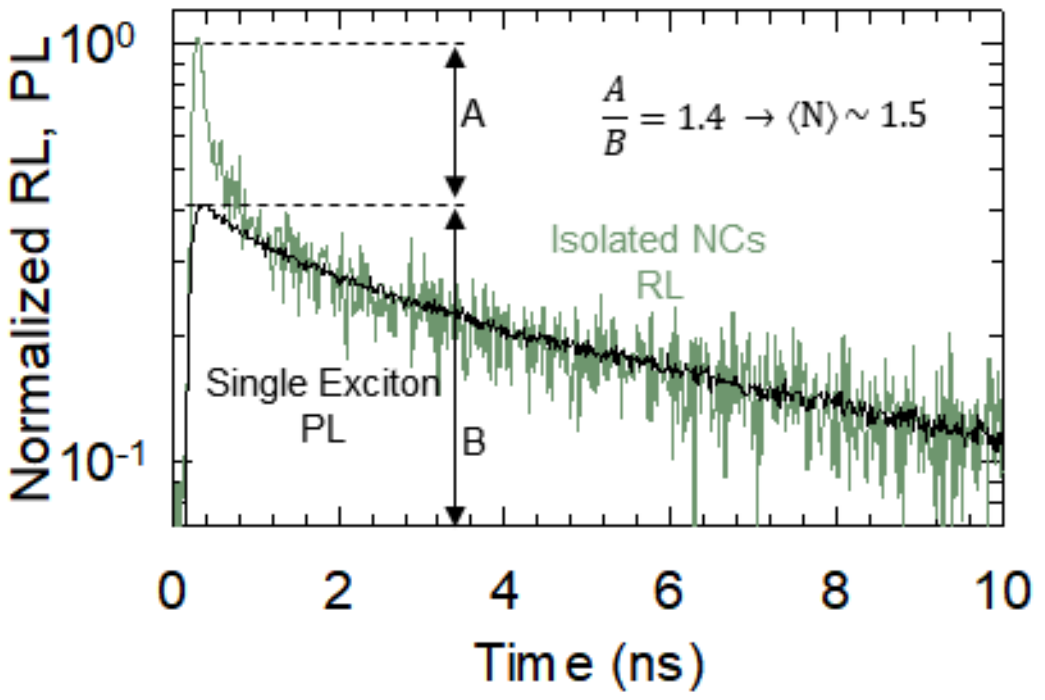

**Supplementary Figure 2:** Normalised RL decay from isolated $CsPbBr_3$ NCs dispersed in an octane solution at room temperature (green), together with the corresponding PL (black), which was excited at a low power to generate only a single excitonic species. The average excitonic population ⟨N⟩ can be inferred from the ratio of A and B, as described in the figure, assuming Poisson statistics of NC initial occupancies at zero-time. The PL is collected within the same setup used for time-resolved RL using a ps-pulsed 3.06eV diode laser.

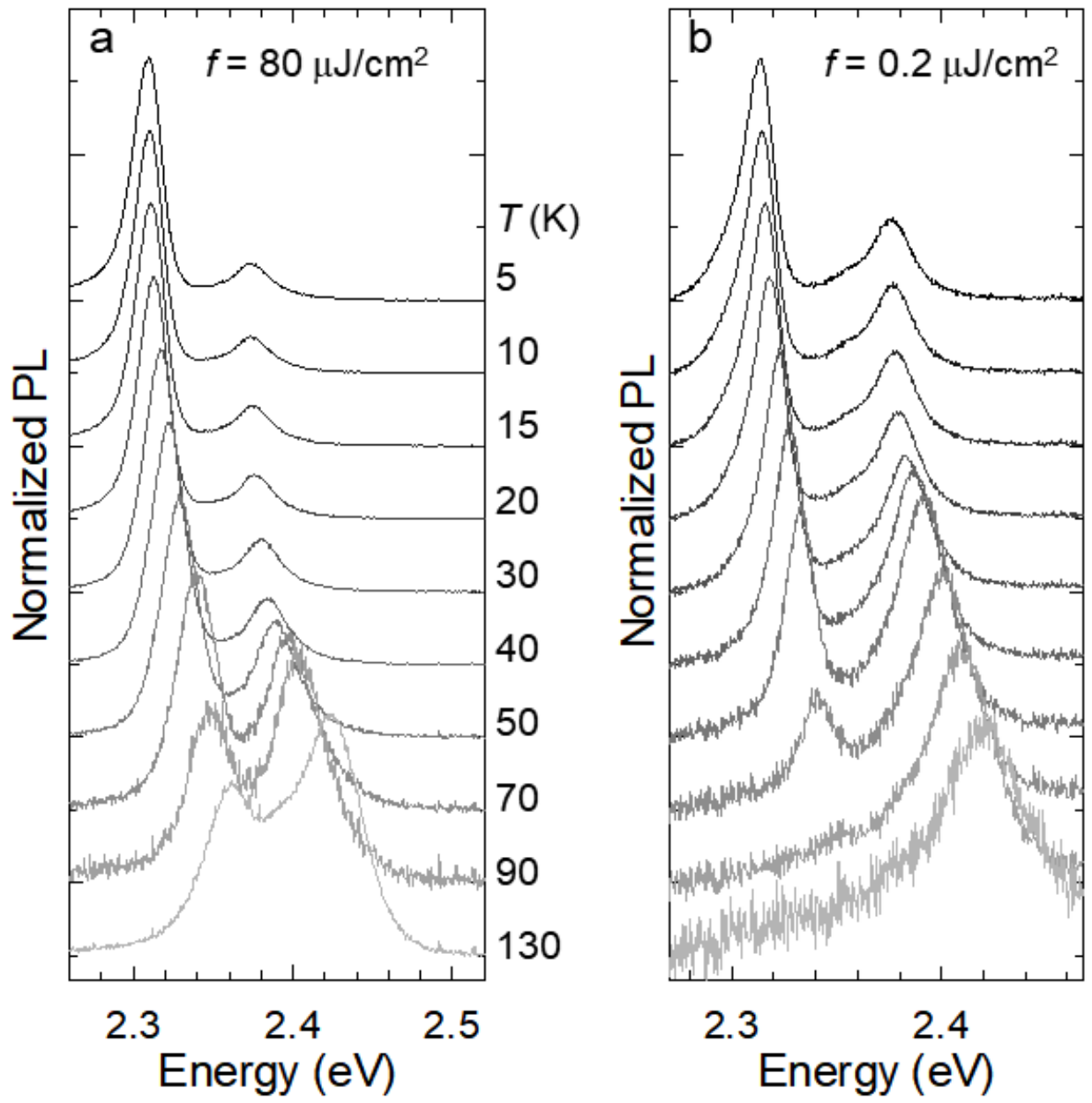

**Supplementary Figure 3:** PL spectra of $CsPbBr_3$ SL at representative temperatures for two optical excitation fluences (*f*), as indicated in figure. The PL spectra in panel 'a' were excited using a 15 kHz pulsed 3.06 eV output from an optical parametric amplifier and measured using a CCD. The PL spectra in panel 'b' were excited using the doubled output of a Ti:Sapphire laser operated at 76 MHz and measured using a streak camera. See the Methods section for details.

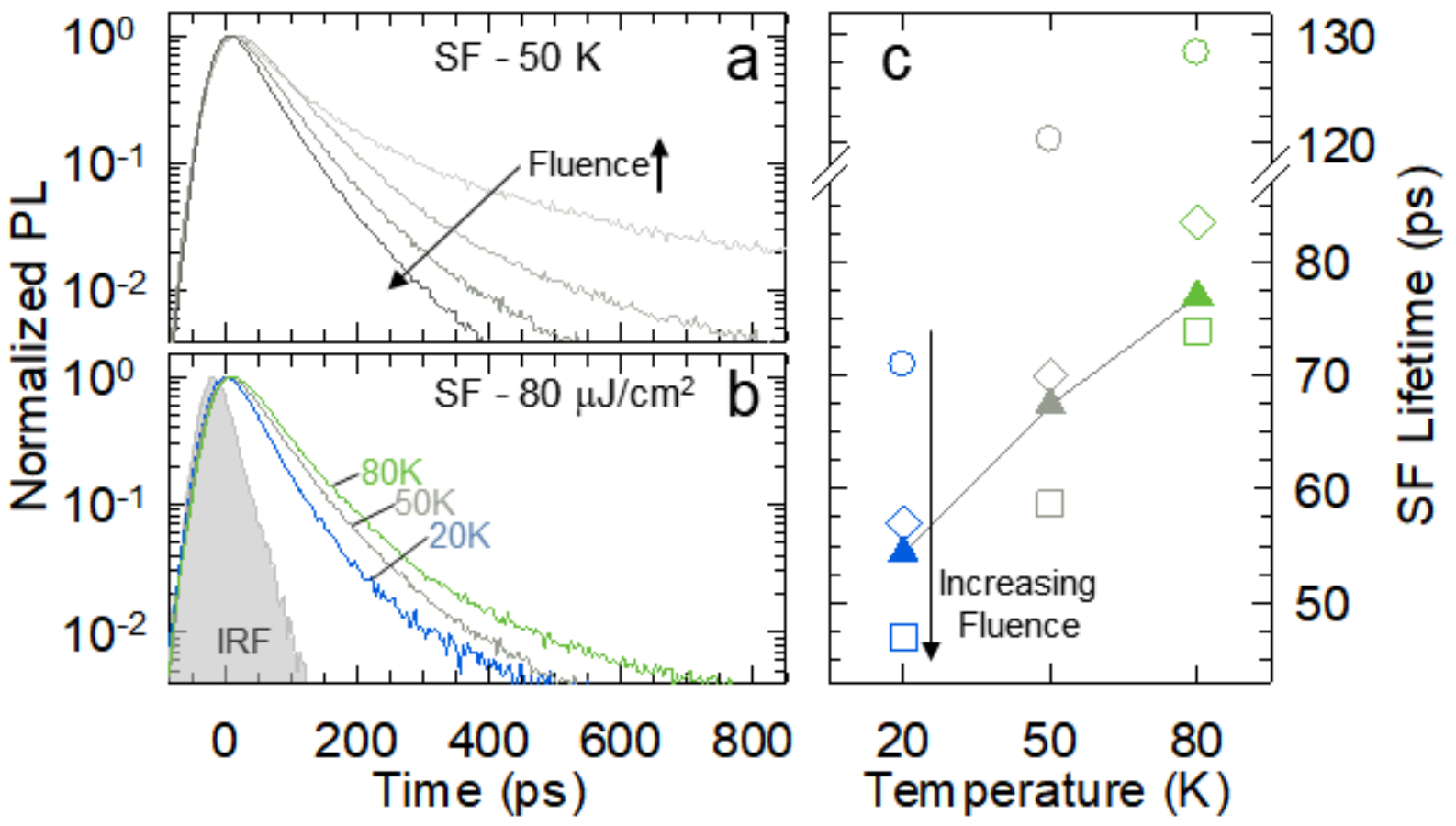

**Supplementary Figure 4: a,** Normalized PL decays of the SF component for $CsPbBr_3$ SLs at 50 K at increasing excitation fluence (2, 20, 80, and 1000 μJ/cm²). **b,** Temperature dependent normalized PL decays of the same SF component under constant excitation fluence (80 μJ/cm²). The shaded area corresponds to the detector response function (IRF). **c,** Effective lifetime of the SF component (defined as the time after which the PL intensity drops by a factor of *e*) as a function of temperature and excitation fluence. The filled symbols correspond to the 80 μJ/cm² excitation fluence.

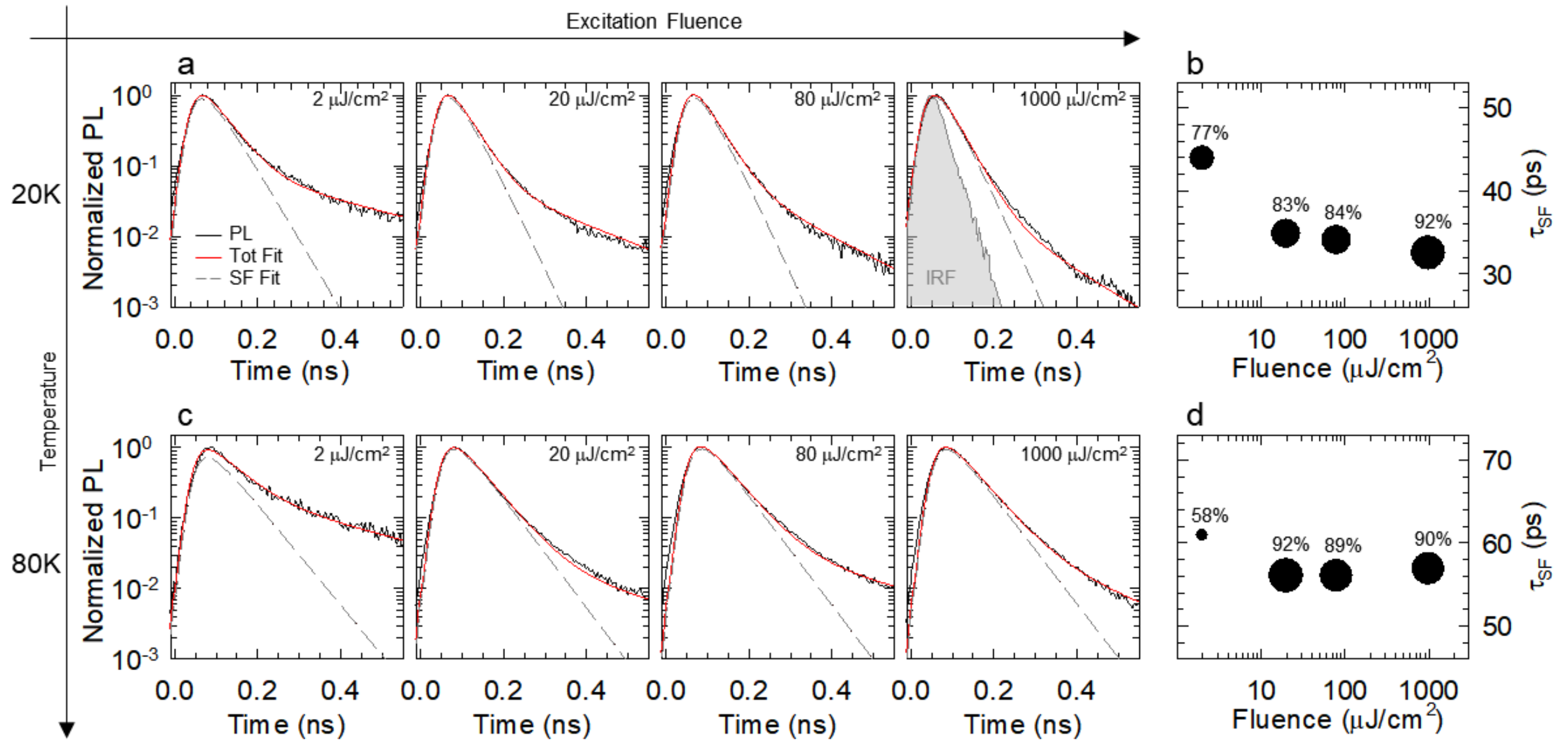

**Supplementary Figure 5: a,** SF decays of $CsPbBr_3$ SLs at 20 K and increasing excitation fluences, as indicated in figure. The red lines correspond to the best fit with a double exponential decay function convoluted with the instrument response function (*vide infra* for details). The dashed lines correspond to the SF component as extracted from the fitting procedure. The PL was excited using a fs-pulsed laser at 3.06 eV operated at 15 kHz. **b,** SF lifetimes extracted from the fitting procedure shown in 'a' as a function of the excitation fluence. The symbol's size scales with the relative intensity of the SF component which is also labelled in the panel for clarity. **c,d,** Same as 'a' and 'b' at 80 K.

The PL decay fitting procedure was carried out in a least-squares fitting approach with the following formula, which accounts for the convolution with the instrument response function (IRF):

$$F(t) = IRF(t) \otimes \left( H(t - t_0) \cdot \left[ \sum_{i=1}^{2} a_i \cdot e^{-t/\tau_i} \right] \right) + C$$

where $t_0$ corresponds to the time start of the emission process, C is the electronic background noise floor, and H is the Heaviside function. The weight of each component was calculated as the integral of each convoluted function over the entire time window.

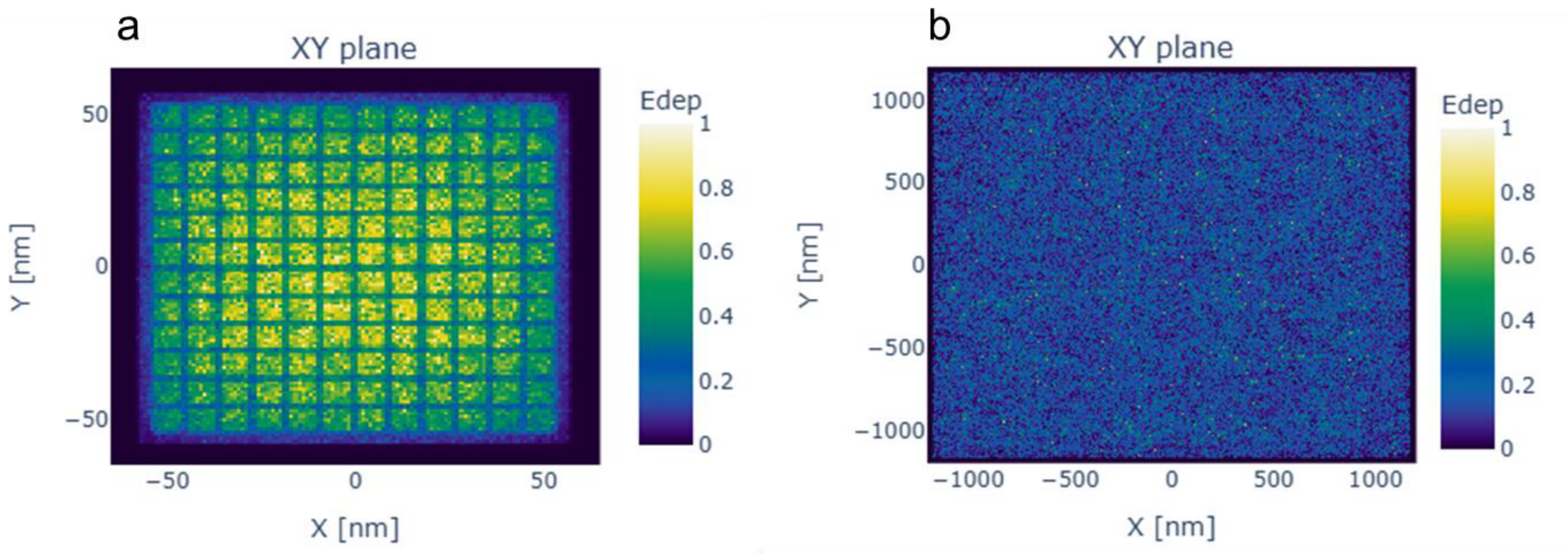

**Supplementary Figure 6:** Orthogonal projections onto the plane XY of the energy deposited (normalized over the maximum deposited) by external X-rays with same energy distribution used for the experimental measurements, in a cubic SL composed of 1728 $CsPbBr_3$ NCs (**a**) and in a flat lattice with a parallelepiped volume of area 5.5 μm$^2$ and thickness of 114 nm composed of 811200 NCs (**b**) regularly arranged and embedded in a surrounding OLAm matrix.

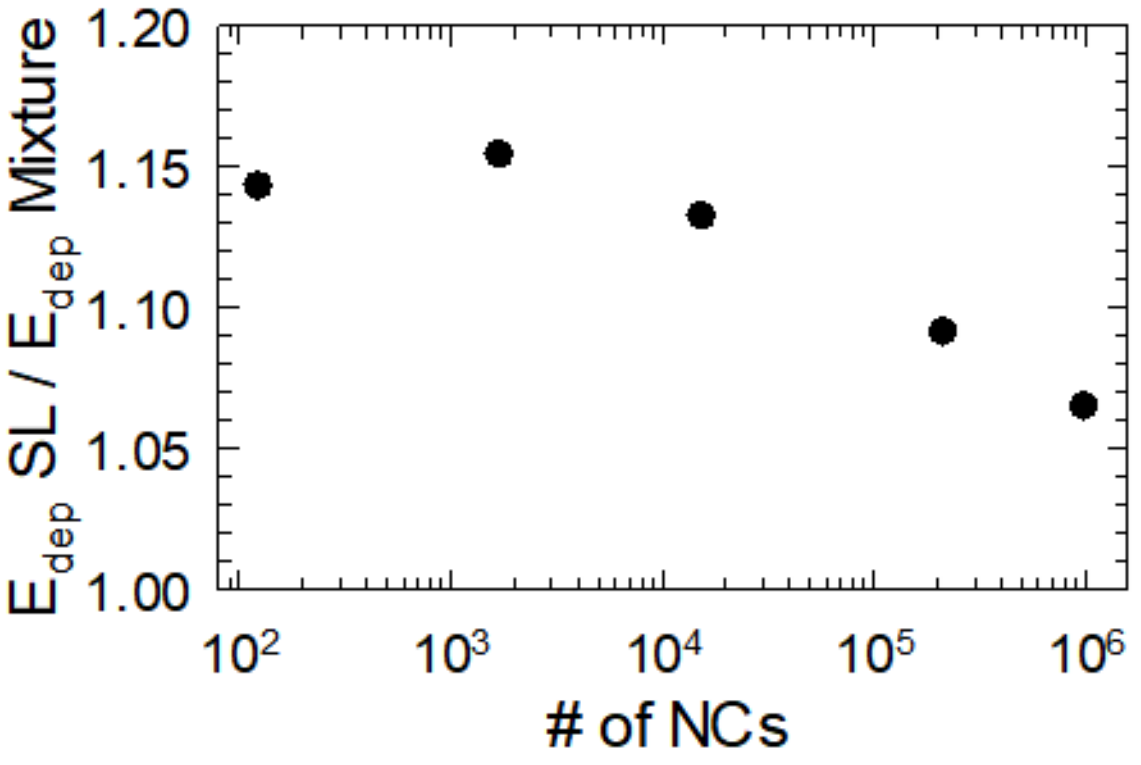

**Supplementary Figure 7:** Ratio of the energy deposited in the SL, modelled as a 3D cubic reticulum, with $CsPbBr_3$ NCs of side 7.6 nm separated by 1.4 nm of OLAm, to the energy released in the same SL, but represented as a single volume made of a mixture of $CsPbBr_3$ and OLAm in various percentages, representing the respective fractions of mass. As the number of NCs increases, the ratio tends to one.